\documentclass[,final,sort&compress]{aipproc}

\layoutstyle{6x9}
   \usepackage{graphicx}
   \usepackage{epstopdf}
  \usepackage{amssymb}

\def\teq#1{$\, #1\,$}                           
\def\fsc{\alpha_{\hbox{\sevenrm f}}}                           
\def\erg{\varepsilon}
\def\sigt{\hbox{$\sigma_{\hbox{\fiverm T}}$}}  
\def\LX{L_{\hbox{\sevenrm X}}}
\def\LXscal{L_{\hbox{\sevenrm X,35}}}
\def\effrad{\epsilon_{\hbox{\sevenrm rad}}}                           
  
\def\rhoGJ{\rho_{\hbox{\fiverm GJ}}}  
\def\Esd{{\dot E}_{\hbox{\fiverm SD}}}  
\def\Gammacyc{\Gamma_{\hbox{\sevenrm cyc}}}   

\def\lambdaC{\lambda_{\hbox{\sevenrm C}}}  
\def\lambdares{\lambda_{\hbox{\eightrm res}}}  
\def\dover#1#2{\hbox{${{\displaystyle#1 \vphantom{(} }\over{
   \displaystyle #2 \vphantom{(} }}$}}                
{\catcode`\@=11                                                  
   \gdef\SchlangeUnter#1#2{\lower2pt\vbox{\baselineskip 0pt\lineskip0pt    
   \ialign{$\m@th#1\hfil##\hfil$\crcr#2\crcr\sim\crcr}}}}           
\def\gtrsim{\mathrel{\mathpalette\SchlangeUnter>}}               
\def\lesssim{\mathrel{\mathpalette\SchlangeUnter<}}
\font\fiverm=cmr5             \font\sevenrm=cmr7     
     \font\eightrm=cmr8

\begin{document}
\newcommand{\vol}[2]{$\,$\bf #1\rm , #2.}                 
\begin{flushright}
\phantom{p}
\vspace{-50pt}
        In Proc. of the Huangshan conference
        ``Astrophysics of Compact Objects,''  (2008)\\
        eds. Y.-F. Yuan, X.-D. Li and D. Lai, (AIP Conf. Proc. 968, New York) p.~93.
\end{flushright} 

\title{Modeling the Non-Thermal X-ray Tail Emission of Anomalous X-ray Pulsars}


\author{Matthew G. Baring}{
  address={Department of Physics and Astronomy, MS 108, 
              Rice University, MS 108, \\
              6100 Main St., Houston, TX 77005, USA.
              {E-mail: \it baring@rice.edu\rm}}
}

\author{Alice K. Harding}{
  address={Gravitational Astrophysics Laboratory,
              NASA Goddard Space Flight Center, Code 663,\\
              Greenbelt, MD 20771, USA.
               {E-Mail: \it harding@twinkie.gsfc.nasa.gov\rm}}
}

\begin{abstract}
The paradigm for Anomalous X-ray Pulsars (AXPs) has evolved recently
with the discovery by INTEGRAL and RXTE of flat, hard X-ray components
in three AXPs.  These non-thermal spectral components differ
dramatically from the steeper quasi-power-law tails seen in the classic
X-ray band in these sources, and can naturally be attributed to activity
in the magnetosphere. Resonant, magnetic Compton upscattering is a
candidate mechanism for generating this new component, since it is very
efficient in the strong fields present near AXP surfaces.  In this
paper, results from an inner magnetospheric model for upscattering of
surface thermal X-rays in AXPs are presented, using a kinetic equation
formalism and employing a QED magnetic scattering cross section.
Characteristically flat and strongly-polarized emission spectra are
produced by non-thermal electrons injected in the emission region.
Spectral results depend strongly on the observer's orientation and the
magnetospheric locale of the scattering, which couple directly to the
angular distributions of photons sampled. Constraints imposed by the
Comptel upper bounds for these AXPs are mentioned.
\end{abstract}

\maketitle


\section{Introduction}
 \label{sec:intro}

A topical focus of the high energy astrophysics of compact objects over
the last decade has been the so-called {\bf magnetars} (Duncan \&
Thompson 1992), constituted by Soft-Gamma Repeaters (SGRs) and Anomalous
X-ray Pulsars (AXPs), whose amassed observational properties have
indicated that they are isolated neutron stars with ultra-strong
magnetic fields. The AXPs, are a group of 6--7 pulsating X-ray sources
with periods around 6-12 seconds.  They are bright, possessing peak
luminosities \teq{\LX \sim 10^{35}\,\rm erg\; s^{-1}}, show no sign of
any companion, are steadily spinning down and have ages \teq{\tau
\lesssim 10^5} years (e.g. Vasisht \& Gotthelf 1997).  Details of the
persistent pulsed X-ray emission for AXPs are discussed in Tiengo et al.
(2002), for XMM observations of 1E 1048.1-5937, and Juett et al. (2002)
and Patel et al. (2003), for the {\it Chandra} spectrum of 4U 0142+61.  
This emission displays both thermal contributions, which have
\teq{kT\sim 0.5-1} keV and so are generally hotter than those in
isolated pulsars, and also non-thermal components with steep spectra
that can be fit by power-laws \teq{dn/dE\propto E^{-s}} of index in the
range \teq{s=2-3.5} (see Perna et al., 2001, for spectral fitting of
ASCA data on AXPs).

The recent detection by the IBIS imager on INTEGRAL, and the PCA and
HEXTE detectors of the Rossi X-ray Timing Explorer, of hard, non-thermal
pulsed tails in three AXPs has provided a new twist to the AXP
phenomenon.  In all of these, the differential spectra above 20 keV are
extremely flat: 1E 1841-045 (Kuiper, Hermsen \& Mendez 2004) has a
power-law energy index of \teq{s=0.94} between around 20 keV and 150
keV, 4U 0142+61  displays an index of \teq{s=0.2} in the 20 keV -- 50
keV band, with a steepening at higher energies implied by the total
DC+pulsed spectrum (Kuiper et al. 2006), and RXS J1708-4009 has
\teq{s=0.88} between 20 keV and 150 keV (Kuiper et al. 2006).  The tails
are much flatter than the non-thermal spectra in the \teq{< 10} keV
band, and do not continue much beyond the IBIS energy window: there are
strongly constraining upper bounds from Comptel observations of these
sources that necessitate a break somewhere in the 150--750 keV band (see
Figures 4, 7 and 10 of Kuiper et al. 2006).

This paper summarizes results from our initial exploration (Baring \&
Harding 2007) of the production of non-thermal X-rays by inverse Compton
heating of soft, atmospheric thermal photons by relativistic electrons,
serving as a model for generating the hard X-ray tails in AXPs.  The
electrons are presumed to be accelerated along either open or closed
field lines with super-Goldreich-Julian densities, perhaps by
electrodynamic potentials, or large scale currents associated with
twists in the magnetic field structure (e.g. Thompson \& Beloborodov
2005). In the strong fields of the inner magnetospheres of AXPs, the
inverse Compton scattering is predominantly resonant at the cyclotron
frequency, with an effective cross section well above the classical
Thomson value. Hence, proximate to the neutron star surface, in regions
bathed intensely by the surface soft X-rays, this process is extremely
efficient for an array of magnetic colatitudes, probably dominating
other processes such as synchrotron and bremsstrahlung radiation that
are employed in the models of Thompson \& Beloborodov (2005) and Heyl 
\& Hernquist (2005).  This prospect motivates the investigation of resonant
inverse Compton models.  Here, the general character of emission spectra
is presented, using collision integral analyses that will set the scene
for future explorations using Monte Carlo simulations.  

\section{Resonant Compton Upscattering in AXPs}

In devising any radiation emission model to describe the non-thermal
X-ray tail luminosity from AXPs, it is necessary to ascertain the
criteria that must be satisfied in order to explain the energetics. 
These were discussed in Baring \& Harding (2007), who determined that
for radiative processes that were electromagnetic in origin, i.e.
involved electrons, the requisite electron densities must be
super-Goldreich-Julian.  This requirement is rather general in nature,
not being constrained to just the resonant Compton upscattering scenario
that is the focus here, but being coupled to a presence of relativistic
electrons moving along {\bf B}, with an abundance that can power the
intense AXP X-ray luminosities, \teq{\LX \gtrsim 10^{35}}erg/sec above
10 keV (Kuiper et al. 2006). The hard X-ray tail luminosities are 2--3
orders of magnitude greater than the classical spin-down luminosity 
\teq{\Esd\sim 8\pi^4 B_p^2R^6/(3P^4c^3)} due to magnetic dipole
radiation torques, where \teq{B_p} is the surface polar field strength,
\teq{P} is the pulsar period and \teq{R} is the stellar radius.

Here we briefly recapitulate the energetics analysis of Baring \&
Harding (2007).  Let \teq{n_e} be the number density of emitting
electrons, \teq{\langle \gamma_e\rangle} be their mean Lorentz factor,
and \teq{\effrad} be the radiative efficiency during their traversal of
the magnetosphere, either along open or closed field lines. Then
\teq{\LX \sim \effrad \langle \gamma_e\rangle m_ec^2 (4\pi n_e R_c^2c)}
if the emission column has a base that is a spherical cap of radius
\teq{R_c < R}. If \teq{R_c\sim 10^6}cm, this yields number densities
\teq{n_e\sim 3\times 10^{17} \LXscal /\effrad \langle \gamma_e\rangle}
cm$^{-3}$ for scaled luminosities \teq{\LXscal \equiv \LX
/10^{35}}erg/sec. Comparing \teq{en_e} to the classic Goldreich-Julian
(1969) density \teq{\rhoGJ =\nabla .\vec{E}/4\pi =
-\vec{\Omega}.\vec{B}/(2\pi c)} for force-free, magnetohydrodynamic
rotators, one can establish the ratio
\begin{equation}
   \dover{en_e}{\vert\rhoGJ\vert} \;\approx\; \dover{4,670}{\effrad \langle \gamma_e\rangle}\;
   \dover{L_{X,35}\, P}{B_{15} R_6^2}\quad ,
 \label{eq:GJ_compare}
\end{equation}
for AXP pulse periods \teq{P} in units of seconds, polar magnetic fields
\teq{B_{15}} in units of \teq{10^{15}}Gauss, and cap radii \teq{R_6} in
units of \teq{10^6}cm.  For \teq{\effrad \langle \gamma_e\rangle\gtrsim
10^3} and \teq{R_6\sim 1}, the requisite density \teq{n_e} is
super-Goldreich-Julian, but not dramatically so.  No choice of radiation
mechanism has been invoked in this line of reasoning.  Yet observe that
electron Lorentz factors of \teq{\gamma_e\gg 10^2-10^3} and their
efficient resonant Compton cooling (i.e. \teq{\effrad\sim 0.01 - 1}) are
readily attained in isolated pulsars with \teq{B\sim 0.1} (e.g. see
Sturner 1995; Harding \& Muslimov 1998; Dyks \& Rudak 2000).  Such
efficiencies should persist into the magnetar regime. To substantiate
this assertion, observe that in the Thomson cooling regime, the resonant
Compton cooling rate \teq{\vert\dot{\gamma}_e\vert \sim (n_s\sigt c)\,
3\pi B^2/(4\fsc \gamma_e \erg_s^2)} can be obtained from Eq.~(17) of
Baring \& Harding (2007), correcting a missing factor of \teq{\fsc
=e^2/(\hbar c)} in the denominator.  The lengthscale \teq{\lambdares
=c/\vert\dot{\gamma}_e\vert} for resonant cooling can then be estimated
from the Planck spectrum photon number density \teq{n_s\sim {\cal
T}^3/\lambdaC^3} at the surface, where the scaled temperature \teq{{\cal
T}=kT/(m_ec^2)} sets \teq{\erg_s\sim 3{\cal T}}, and \teq{\lambdaC=
\hbar /m_ec} is the electron Compton wavelength over \teq{2\pi}.  The
result is \teq{\lambdares \sim (\lambdaC/\fsc ) \, \gamma_e /({\cal T}
B^2)}, lengthens considerably at high altitudes due to dilution of the
soft photon density.  For \teq{B\sim 0.1}, \teq{\gamma_e\sim 10^3} and
\teq{{\cal T}\sim 3\times 10^{-3}}, \teq{\lambdares} is much less than
the stellar radius, a result that is clearly not modified by
relativistic quantum corrections when \teq{B\gtrsim 1}.

\begin{figure}
   \centerline{
   \includegraphics[width=0.53\textwidth]{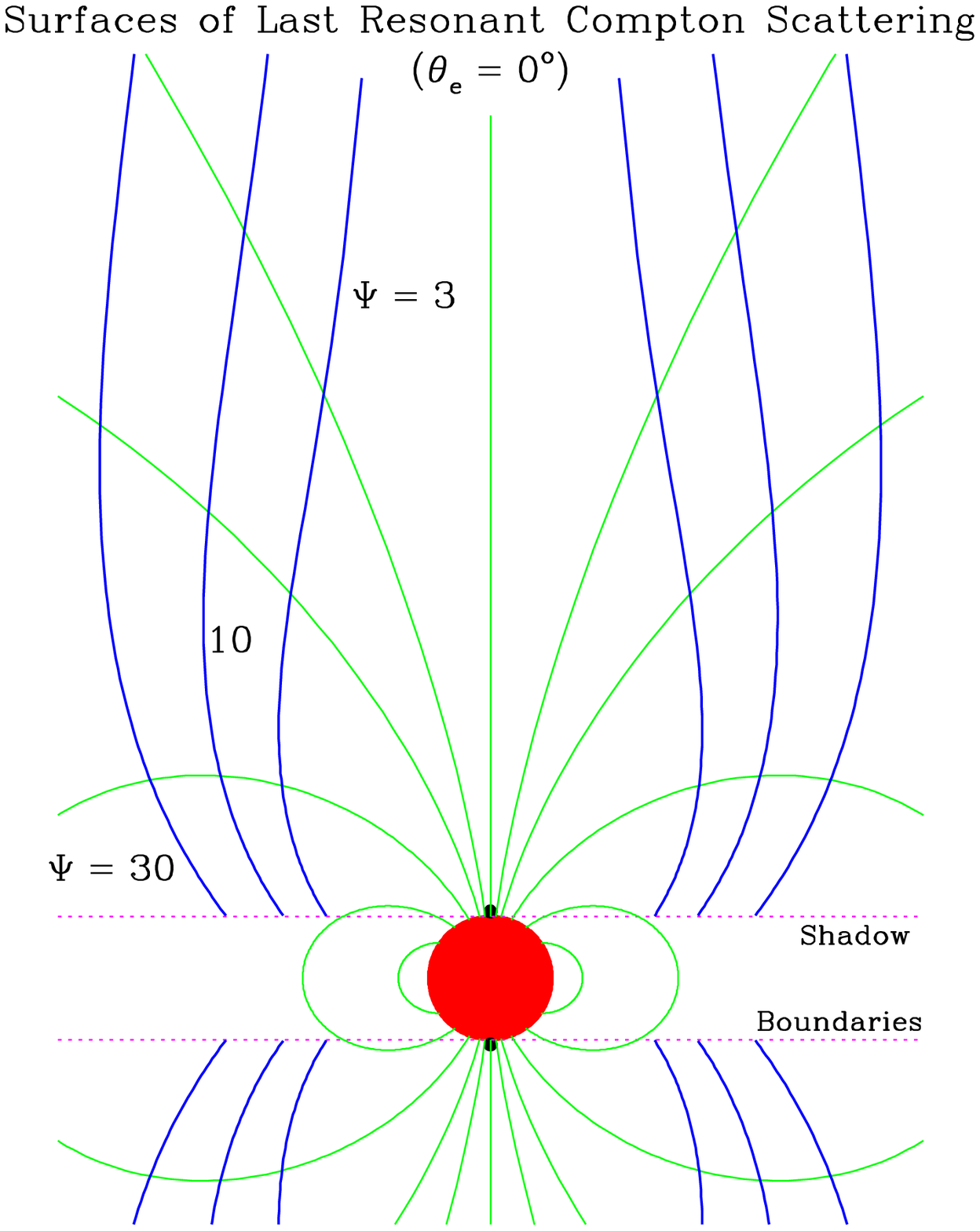}
   \hskip 10pt
   \includegraphics[width=0.53\textwidth]{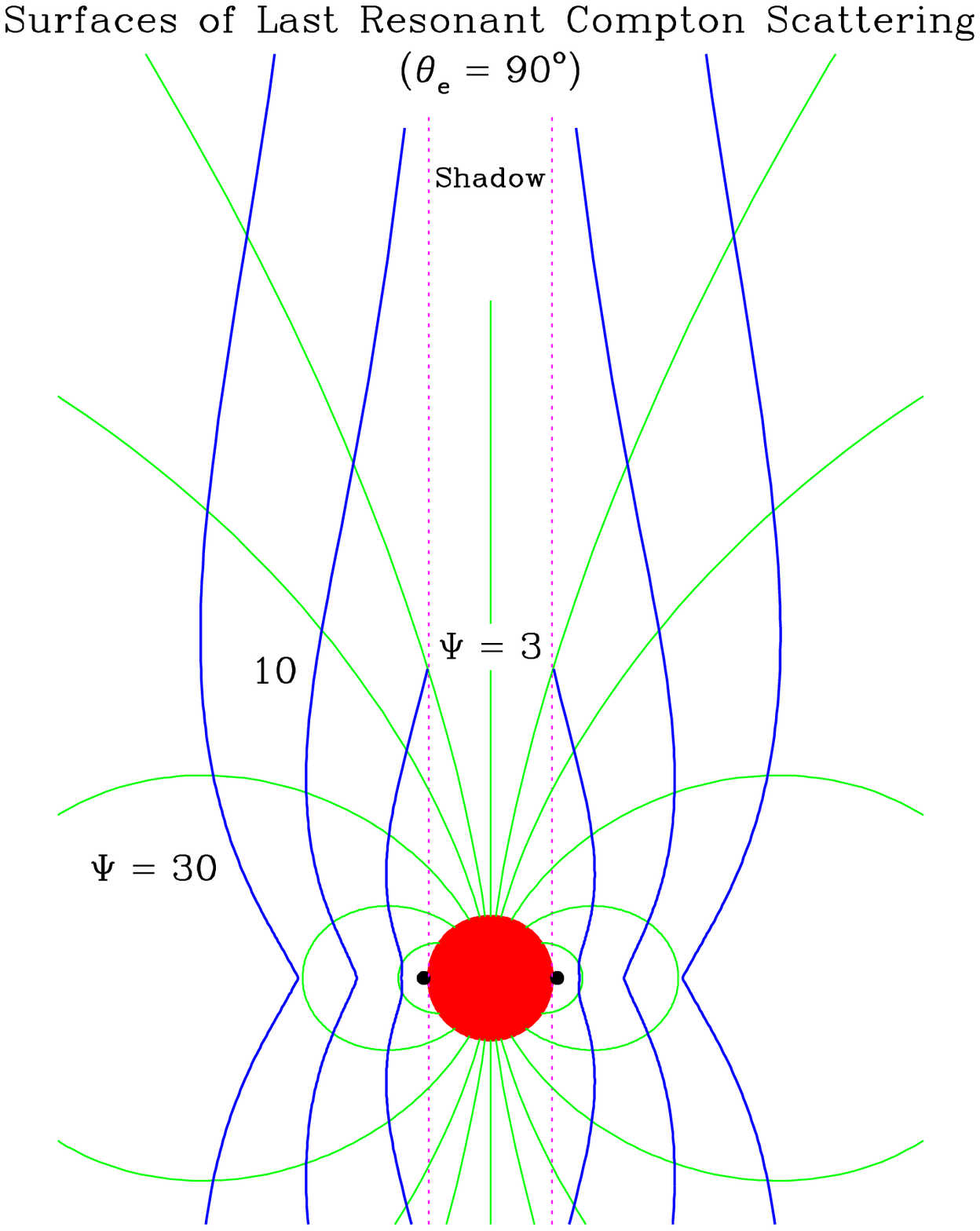}
   \vspace{-100pt}}
   \caption{Contours in a section of a pulsar magnetosphere that depict cross
	sections of the surfaces of last resonant scattering by
	ultra-relativistic electons, i.e. the maximal extent of the Compton
	resonasphere.  The heavyweight blue contours are computed for different
	values of the resonance parameter \teq{\Psi} defined in
	Eq.~(\ref{eq:surface_locus}), and at extremely high altitudes
	asymptotically approach the magnetic axis (central vertical line). The
	filled red circles denote the neutron star, whose radius \teq{R}
	establishes the linear spatial scale for the Figure. The cases
	illustrated are for photons emanating from the polar axis (i.e.
	\teq{\theta_e=0^{\circ}}, left panel) and from the equator (i.e.
	\teq{\theta_e=90^{\circ}} right panel), denoted by the small black dots,
	for which the neutron star shadow regions are demarcated by the light
	dotted boundaries, and only the surfaces (azimuthally-symmetric about
	the magnetic axis) are accessible to resonant Compton interactions.
       }
 \label{fig:resonasphere} 
\end{figure}

By necessity, the locales where Compton interactions sample the
cyclotron resonance are confined to the lower altitudes in an AXP
magnetosphere, where the field is sufficiently high. This is controlled
primarily by the scattering kinematics, which dictates a coupling
between the energies \teq{\erg_{\gamma} m_ec^2} and \teq{\gamma_e
m_ec^2} of colliding X-ray photons and electrons, respectively, and the
local angle \teq{\theta_{\gamma}} of the interacting photon to the
magnetic field lines.  The cyclotron fundamental is sampled when
\begin{equation}
   \gamma_e\erg_{\gamma} (1-\cos\theta_{\gamma}) \;\approx\; B\quad , \quad 
   \hbox{for}\; \gamma_e\;\gg\; 1\quad .
 \label{eq:res_kinematics}
\end{equation}
For X-ray photons emanating from a single point on the stellar surface,
Baring \& Harding (2007) computed the zones of influence of the resonant
Compton process for dipole field geometry in a flat spacetime. Assuming
that the X-rays propagate with no azimuthal component to their momenta,
the resonance criterion is satisfied on a surface that is azimuthally
symmetric about the magnetic field axis, for slow rotators.  For
outward-going electrons, the locus of the projection of this surface
onto a plane intersecting the magnetic axis was found to be
\begin{equation}
   \chi^3\; =\; \Psi\; \dover{\sqrt{1+3\cos^2\theta}}{
       1-\cos\theta_{\gamma}}\quad , \quad
    \Psi\; =\; \dover{B_p}{2\gamma_e\erg_{\gamma}}\quad ,
 \label{eq:surface_locus}
\end{equation}
where \teq{\chi=r/R} is the altitude scaled in units of the neutron star
radius \teq{R}, and \teq{\theta} is the magnetic colatitude of the point
of scattering.  \teq{B_p} is the surface polar field strength in units
of \teq{B_{\rm cr}=4.413\times 10^{13}}Gauss. Geometry determines the
function, \teq{\theta_{\gamma} = \theta_{\gamma} (\theta_e,\, \theta )},
which is given in Eq.~(5) of Baring \& Harding (2007), where
\teq{\theta_e} is the colatitude of the surface emission point. Here
\teq{\Psi} is the key parameter that scales the altitude of the locale
of resonant interaction, and typically falls in the range \teq{1-10^3}
for magnetars when \teq{\gamma_e\sim 10^2 - 10^4}. For the broadly
representative cases of soft photons emitted from the surface pole
(\teq{\theta_e=0}) and magnetic equator (\teq{\theta_e=90^{\circ}}), the
surfaces of resonant scattering for different \teq{\Psi} are illustrated
in Fig.~\ref{fig:resonasphere}. Shadows of the emission points are also
indicated to mark propagation exclusion zones for the chosen emission
colatitudes.

%

It is evident from the Figure that the altitude of resonance is much
lower for equatorial emission cases (comparing left and right panels),
and also in the equatorial regions when compared with polar locales
(within each panel).  At small colatitudes above the magnetic pole,
\teq{\theta_{\gamma}} is necessarily small, pushing the resonant surface
to very high altitudes where the field is much lower. In equatorial
interaction locales, which are preferentially sampled for
quasi-equatorial emission colatitudes, the photons tend to travel more
across field lines in the observer's frame, and so access the resonance
in regions of higher field strength, thereby reducing the altitudes
where the resonance is sampled.  These are manifestations of the
correlation between \teq{\theta_{\gamma}} and \teq{B} (for fixed
\teq{\erg_{\gamma}} and \teq{\gamma_e}) evinced in
Eq.~(\ref{eq:res_kinematics}). For cases  \teq{0 < \theta_e <
90^{\circ}} (not depicted), the contours are morphologically similar,
though they incur significant deviations from those in
Fig.~\ref{fig:resonasphere}.  Clearly, by sampling different emission
colatitudes \teq{\theta_e} these surfaces are smeared out into annular
volumes.  Observe that introducing an azimuthal component to the photon
momentum tends to increase propagation across the field, i.e. raising
\teq{\theta_{\gamma}}, so that the resonance is accessed at lower
altitudes and higher field locales.  Hence, loci like those depicted in
Fig.~\ref{fig:resonasphere} actually represent the outermost extent of
resonant interaction, and so are {\it surfaces of last resonant
scattering}, i.e. the outer boundaries to the {\bf Compton
resonasphere}.  It is evident that, for the majority of closed field
lines for long period AXPs, this resonasphere is confined to within a
few stellar radii of the surface. 

\subsection{Resonant Compton Upscattering Spectra}

Collision integral calculations for upscattering spectra from resonant
Compton interactions are routinely obtained for uniform magnetic fields.
In the AXP problem, the scalelength \teq{\lambdares} for the interaction
is often much shorter than the scale \teq{R} of the field
divergence/gradient, so such uniform {\bf B} computations are reasonably
informative. Let \teq{n_{\gamma}} be the number density of photons
resulting from the resonant upscattering process. For inverse Compton
scattering, an expression for the spectrum of photon production
\teq{dn_{\gamma}/(dt\, d\erg_f\, d\mu_f) }, differential in the photon's
post-scattering laboratory frame quantities \teq{\erg_f} and
\teq{\mu_f=\cos\Theta_f}, was presented in Eqs.~(A7)--(A9) of Ho and
Epstein (1989), valid for general scattering scenarios.  The
dimensionless pre- and post-scattering photon energies (i.e. scaled by
\teq{m_ec^2}) in the observer's frame (OF) are \teq{\erg_i} and
\teq{\erg_f}, respectively, and the corresponding angles of these
photons with respect to the OF electron velocity vector
\teq{-\vec{\beta}_e} (i.e. field direction) are \teq{\Theta_i} and
\teq{\Theta_f}, respectively. The result for the spectrum can be
integrated over \teq{\mu_f} and then written as (detailed in Baring \&
Harding, 2007)
\begin{equation}
   \dover{dn_{\gamma}}{dt\, d\erg_f} \, =\,  \dover{n_e n_s\, c}{\mu_+-\mu_-}
   \int_{-1}^{1}d\mu_f   \int_{\mu_-}^{\mu_+}d\mu_i \; \delta \big\lbrack\omega_f -\omega'(\omega_i,\,\theta_f)\, \bigl\rbrack\,
   \dover{1+\beta_e\mu_i}{\gamma_e(1+\beta_e\mu_f)}\,
   \dover{d\sigma}{d(\cos\theta_f) }\;\; .
 \label{eq:scatt_spec}
\end{equation}
Here, the notation \teq{\mu_i=\cos\Theta_i} and \teq{\mu_f=\cos\Theta_f}
is used for compactness, \teq{n_e} and \teq{n_s} are the number
densities of relativistic electrons and soft photons, respectively.  
For simplicity, the incident photons are assumed to be monoenergetic and
to possess a uniform distribution of angle cosines \teq{\mu_i} in some
range \teq{\mu_-\leq\mu_i\leq \mu_+}. Note also that since
\teq{\gamma_e\gg 1}, the incident photon angle with respect to {\bf B}
in the electron rest frame (ERF) is \teq{\theta_i\approx 0}, while the
ERF angle \teq{\theta_f} of the scattered photon relative to the field
can take on a range of values. The function
\teq{\omega'(\omega_i,\,\theta_f)}, which appears in the \teq{\delta}
function in Eq.~(\ref{eq:scatt_spec}), encapsulates the electron rest
frame scattering kinematics, as detailed in Baring \& Harding (2007). 
The differential cross section, \teq{d\sigma/d(\cos \theta_f)},
appearing in Eq.~(\ref{eq:scatt_spec}) is evaluated in the ERF, and is
taken from Eq.~(23) of Gonthier et al. (2000); it incorporates
relativistic QED physics that is applicable for arbitrary field
strengths.  Specialized to the case of scatterings that leave the
electron in the ground state, the zeroth Landau level that it originates
from, its specific form is summarized in Baring \& Harding (2007).  More
general results for the fully relativistic, quantum cross section for
resonant Compton scattering can be found in Herold (1979), Daugherty \&
Harding (1986), and Bussard, Alexander \& M\'esz\'aros (1986). These
extend earlier non-relativistic quantum mechanical formulations such as
in Canuto, Lodenquai \& Ruderman (1971), and Blandford \& Scharlemann
(1976).

%
\begin{figure}[ht]
   \centering
   \includegraphics[width=0.63\textwidth]{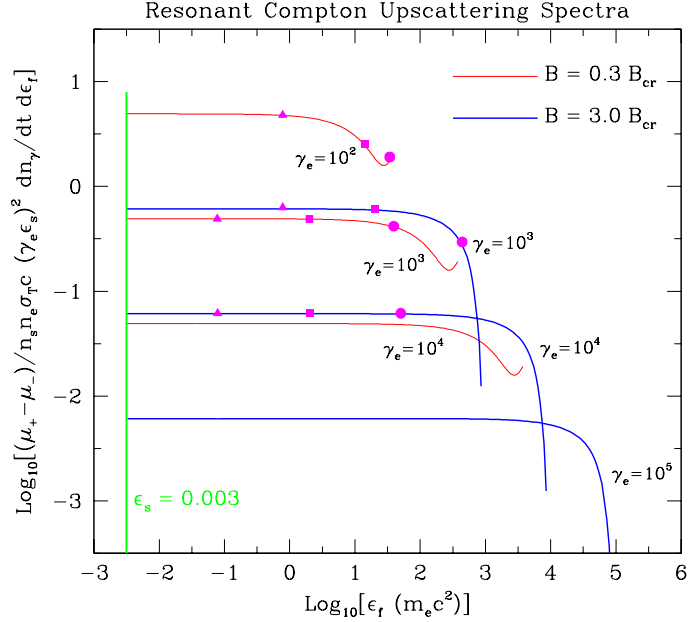}
   \caption{Resonant Compton upscattering spectra (scaled, unpolarized) 
   such as might be sampled in the magnetosphere of an AXP, for different 
   relativistic electron Lorentz factors \teq{\gamma_e}, as labelled. 
   The emergent photon energy \teq{\erg_f} is scaled in terms of 
   \teq{m_ec^2}.  The chosen magnetic field strengths of \teq{B=3 B_{\rm cr}} 
   (heavyweight, blue) and \teq{B=0.3 B_{\rm cr}} (lighter weight, red) 
   correspond to different altitudes and perhaps colatitudes.  
   Results are depicted for seed photons of energy \teq{\erg_s=0.003}
   (marked by the green vertical line), typical of thermal X-rays emanating 
   from AXP surfaces; downscattering resonant emission at 
   \teq{\erg_f < \erg_s} was not exhibited.   Specific emergent angles of the 
   emission in the observer's frame, with respect to the magnetic field
   direction, are indicated by the filled magenta symbols for four of the spectra, with 
   triangles denoting \teq{\Theta_f=5^{\circ}}, squares corresponding to 
   \teq{\Theta_f=1^{\circ}}, and circles representing \teq{\Theta_f=0.2^{\circ}}.
   }
 \label{fig:upscatt_spectra} 
\end{figure}

The relativistic Compton cross section \teq{\sigma} is strongly peaked
at the cyclotron fundamental (see Fig.~2 of Gonthier et al. 2000) due to
the appearance of resonant denominators in the S-matrices. These
generate a Lorentz profile factor \teq{1/[(\omega_i-B)^2 + (\Gamma
/2)^2]} in \teq{d\sigma/d(\cos\theta_f)}, where \teq{\Gamma =
\Gammacyc\, E_1} and \teq{E_1=\sqrt{1+2B+\omega_i^2}} is the energy of
the intermediate electron.  Here \teq{\Gammacyc\ll B} is the
dimensionless cyclotron decay rate from the first Landau level (with
electron momentum \teq{p_z=\omega_i} parallel to {\bf B}), signifying
that the intermediate electron states become effectively real at the
cyclotron resonance, and possess a finite decay lifetime.  A form for
the decay rate \teq{\Gammacyc}, can be found in Eqs.~(13) or (23) of
Baring, Gonthier \& Harding (2005; see also Latal 1986; Harding \& Lai
2006). For \teq{B\ll 1}, \teq{\Gammacyc\approx 4\fsc B^2/3}, while for
\teq{B\gg 1}, quantum and recoil effects generate \teq{\Gammacyc\approx
(\fsc/e) \sqrt{B/2}}.  Note that Baring \& Harding (2007) adopted the
{\it ansatz} \teq{\Gamma\to\Gammacyc} appropriate for resonant Thomson
scattering; here this is updated to incorporate relativistic corrections
to the resonance width following Harding \& Daugherty (1991), amounting
to a spectral normalization correction by a factor of
\teq{1/\sqrt{1+2B}} for all \teq{\erg_f}.

Representative spectral forms are depicted in
Fig.~\ref{fig:upscatt_spectra}, for the situation where emergent
polarizations are not observed (see Baring \& Harding 2007, for
polarization characteristics).  Because of the narrowness of the
resonance, non-resonant scattering contributions were omitted when
generating the curves; these contributions produce steep wings to the
spectra at the uppermost and lowermost energies, and a slight bolstering
of the flat portion.  Generally they contribute significantly only when
access to the resonance is kinematically forbidden, i.e. outside the
Compton resonaspheres illustrated in Fig.~\ref{fig:resonasphere}. The
resonant restriction kinematically limits the emergent photon energies
\teq{\erg_f} to
\begin{equation}
   \gamma_e(1-\beta_e)\, B \;\leq\;\erg_f\;\leq\; 
   \dover{\gamma_e(1+\beta_e)\, B}{1+2B} \quad ,
 \label{eq:kinematic_range}
\end{equation}
a range that generally extends below the thermal photon seed energy
\teq{\erg_s}.  For the \teq{B=0.3} case in
Fig.~\ref{fig:upscatt_spectra}, a quasi-Thomson regime, the spectra are
characteristically flat (e.g. see Dermer 1990; Baring 1994; Liu et al.
2006) for most \teq{\erg_f}, indicative of the kinematic sampling of the
resonance in the integrations over soft photon angles \teq{\Theta_i}.
Since \teq{\Gamma\ll B} in general, the normalization of this flat
portion scales as \teq{dn_{\gamma}/(dt\, d\erg_f) \propto 1/\Gamma}.
Only at the highest energies does the spectrum begin to deviate from
flat (i.e. horizontal) behavior, and this domain corresponds to
significant scattering angles in the ERF, i.e. cosines
\teq{1-\cos\theta_f} not much less than unity. Then the mathematical
form of the differential cross section becomes influential in
determining the spectral shape, as discussed in Baring \& Harding
(2007). In AXPs, the \teq{B=0.3} case best represents higher altitude
locales for the resonasphere, such as at smaller colatitudes near the
polar axis. Fig.~\ref{fig:upscatt_spectra} also exhibits spectra for
\teq{B=3}, a case more typical of equatorial resonance locales.  Then
the flat spectrum still appears at energies \teq{\erg_s\leq\erg_f \ll
\gamma_e(1+\beta_e)\, B/(1+2B)},  when again \teq{\cos\theta_f\approx
1}.  Yet the \teq{B=3} curves in the Figure display more prominent
reductions at the uppermost energies \teq{\erg_f\sim
\gamma_e(1+\beta_e)\, B/(1+2 B)} due to the sampling of \teq{1/2 \leq
\omega_f \ll \omega_i} values in the ERF that correspond to strong
electron recoil effects.

Photons emitted in these uppermost energies have \teq{1-\cos\theta_f\sim
1} in the ERF and are highly beamed along the field in the observer's
frame.  This is an important property that is highlighted via the filled
symbols in Fig.~\ref{fig:upscatt_spectra}.  The intense beaming of
radiation along {\bf B} in the OF, and the profound correlation of the
angle of emission \teq{\Theta_f} with the emergent photon energy
\teq{\erg_f}, are both consequences of scattering kinematics in the
resonance.  In the resonant case here, the \teq{\gamma_e\gg 1} regime
dictates that most of the emission is collimated to within
\teq{5^{\circ}} of the field direction, and rapidly becomes beamed to
within \teq{0.2^{\circ}} as the final photon energy increases towards
its maximum. This kinematic characteristic guarantees that spectral
formation in Compton upscattering models is extremely sensitive to the
observer's viewing orientation in relation to the magnetospheric
geometry.  This suggests powerful geometrical probes of AXP emission
regions if pulse-phase spectroscopy is achievable in future generation
observatories.

\vskip -6pt
\section{Conclusion}
 \label{sec:conclusion}

This paper has summarized some essentials of the resonant Compton
upscattering model for hard X-ray tail emission in AXPs, as developed in
Baring \& Harding (2007). The spectra exhibited in
Fig.~\ref{fig:upscatt_spectra} are considerably flatter than the hard
X-ray tails (\teq{\sim \erg_f^{-\alpha}} for \teq{\alpha\sim 0.2 -1})
seen in the AXPs, and extend to GLAST-band energies much higher than can
be permitted (i.e. around 750 keV) by the Comptel upper bounds to these
sources, unless emergent angles \teq{\Theta_f} to {\bf B} exceed around
\teq{1^{\circ}}. They represent a preliminary indication of how flat the
resonant scattering process can render the spectrum, which can readily
be steepened by spatial distribution of electron injection, significant
and unavoidable cooling, and also non-resonant contributions.  What an
observer detects will depend critically on his/her viewing perspective
and the magnetospheric locale of the scattering. The one-to-one
kinematic coupling between \teq{\erg_f} and \teq{\mu_f} implies that the
highest energy photons are beamed strongly along the local field
direction.  This may or may not be sampled by an instantaneous
observation at a given rotational phase. Realistically, for many pulse
phases, angles corresponding to \teq{\mu_f <1} will be predominant,
lowering the value of \teq{\erg_f}.  How low is presently unclear, and
remains to be explored via a model with full magnetospheric geometry,
the obvious next development in this research program.


\begin{theacknowledgments}
	We thank Lucien Kuiper and Wim Hermsen for discussions concerning
	the INTEGRAL/RXTE data on the hard X-ray tails in Anomalous X-ray
	Pulsars.  Support for this research was provided by NASA via
	Grant Nos.~NNG05GK29G and~NNX06AI32G, and the 
	National Science Foundation through Grant No.~AST06-07651.
\end{theacknowledgments}



\end{document}